\documentclass[aps, pre, twocolumn, superscriptaddress]{revtex4-1}

\usepackage[english]{babel}
\usepackage{hyperref}
\usepackage{bbold}

\usepackage[pdftex]{graphicx}
\usepackage{color}
\usepackage[toc,page]{appendix}

\usepackage{amsmath}
\usepackage{amssymb}
\usepackage{mathtools}
\usepackage{braket}

\usepackage[sort&compress]{natbib}

\newcommand{\fig}[1]{Figure~\ref{fig:#1}}
\newcommand{\Fig}[1]{Figure~\ref{fig:#1}}

\newcommand{\eq}[1]{Eq.~(\ref{eq:#1})}

\newcommand{\I}{\mathrm{i}}

\newcommand{\eps}{\epsilon}

\begin{document}

\title{Robustness of optimal transport in disordered interacting many-body networks}

\author{Adrian Ortega}
\email{adrianortega@fis.unam.mx}
\affiliation{Instituto de Ciencias F\'isicas, Universidad Nacional Aut\'onoma de M\'exico, 62210 Cuernavaca, M\'exico}

\author{Thomas Stegmann}
\email{stegmann@icf.unam.mx}
\affiliation{Instituto de Ciencias F\'isicas, Universidad Nacional Aut\'onoma de M\'exico, 62210 Cuernavaca, M\'exico}

\author{Luis Benet}
\email{benet@fis.unam.mx}
\affiliation{Instituto de Ciencias F\'isicas, Universidad Nacional Aut\'onoma de M\'exico, 62210 Cuernavaca, M\'exico}
\affiliation{Centro Internacional de Ciencias, 62210 Cuernavaca, M\'exico}

\date{\today}


\begin{abstract}
  The robustness of quantum transport under various perturbations is analyzed in disordered
  interacting many-body systems, which are constructed from the embedded Gaussian random matrix
  ensembles (EGEs). The transport efficiency can be enhanced drastically, if centrosymmetry (csEGE)
  is imposed. When the csEGE is perturbed with an ordinary EGE, the transport efficiency in the
  optimal cases is reduced significantly, while in the suboptimal cases the changes are less
  pronounced. Qualitatively the same behavior is observed, when parity and centrosymmetry are broken
  by block perturbations. Analyzing the influence of the environment coupling, optimal transport is
  observed at a certain coupling strength, while too weak and too strong coupling reduce the
  transport. Taking into account the effects of decoherence, in the EGE the transport efficiency
  approaches its maximum at a finite nonzero decoherence strength (environment-assisted
  transport). In the csEGE the efficiency decays monotonically with the decoherence but is always
  larger than in the EGE.
\end{abstract}

\maketitle

\section{Introduction and motivation}
\label{sec:Introduction}

Quantum efficiency assesses the transport probability of particles or excitations across a quantum
system \cite{tor_pre}. A prominent example for these systems is given by photosynthetic
biomolecules, where highly efficient transport can be observed \cite{1997PhysToday, tor_pre,
  lee_science, biolandreas, Scholes2017}. This system can be modeled by a network of sites and bonds
\cite{Adolphs20062778}, or more abstractly, by means of disordered random networks
\cite{2013PRL-walschaers}. In general, random disorder hinders the transport due to Anderson
localization. Hence, it is necessary to identify structural elements, which provide efficient
quantum transport in the presence of disorder. It has been demonstrated that a specific symmetry in
the Hamiltonian, called centrosymmetry, improves significantly the overall transport across the
network \cite{2013PRL-walschaers, 2015PRE-walschaers, Walschaers2017, Walschaers2017a}. Recently,
these studies have been extended to interacting disordered networks, modeled by embedded Gaussian
ensembles (EGEs) \cite{2001AnnPhys-BRW, kotabook, Vyas2018} and their centrosymmetric version
(csEGEs) \cite{2015AdrManBen, PhysRevE.94.042102}. The many-body interactions are reflected by the
correlations in these networks. It has been shown that centrosymmetry induces additional strong
correlations in these systems that enhance drastically the transport \cite{2015AdrManBen,
  PhysRevE.94.042102, OrtegaPhDThesis2017}. In particular, it has been found that in almost filled
systems with a rank of interaction $k \sim n/2$, where $n$ is the number of particles, high quantum
efficiency is observed in almost all random realizations. While at this point it is unclear whether
centrosymmetry is present in photosynthetic biomolecules, it has been implemented in the laboratory
\cite{nikojex}, and it is a necessary ingredient for efficient transfer of quantum states
\cite{akay}, which can be used for the state transfer between quantum processors
\cite{christandl_prl, christandl_pra}.

The aim of this paper is to investigate the robustness of the quantum efficiency in disordered
many-body networks under various perturbations. Starting with a centrosymmetric system, we determine
how noncentrosymmetric perturbations affect the quantum efficiency. Centrosymmetry is essentially
parity and correlations among two sectors of different parity \cite{OrtegaPhDThesis2017} and
generates a block structure in the Hamiltonian \cite{cantoni}. Hence, we analyze the effect of
perturbations that mix sectors with different parity as well as perturbations that mix different
block sectors in the Hamiltonian. We will also investigate how the transport is affected by the
coupling strength of the environment through which the excitations are injected and extracted. The
transport in disordered networks can not only be enhanced by centrosymmetry but also by means of
decoherence \cite{plenio_dephtransp, rebentrost_dephtransp, Roberto_JPC2014, PhysRevE.95.022122},
which is certainly present to some degree in biomolecules at room temperature. Therefore, we study
the effect of decoherence on the transport in disordered networks with and without
centrosymmetry. This will allow us to analyze the interplay between both, centrosymmetry and
decoherence.

\section{Model and methods}
\label{sec:ModelMethods}

\subsection{Embedded random matrix ensemble for disordered interacting systems}
\label{sec:EGE}

In this Section, we introduce the fermionic embedded Gaussian ensemble \cite{2001AnnPhys-BRW,
  kotabook, Vyas2018}, which is used as a tool to describe the statistical properties of interacting
quantum many-body systems. This ensemble has found broad applications in nuclear physics, quantum
information, and quantum transport; see \cite{kotareview}.

This ensembles is constructed in the following way \cite{2001AnnPhys-BRW, 2003JPA-BW, kotabook,
  Vyas2018}. We consider a quantum system of $n$ interacting fermions distributed over
$l$ single-particle states. As we are interested in finite quantum systems, we choose typically low
values for the single-particle number $l$. Furthermore $1\leq n\leq l$ in agreement with Pauli's
exclusion principle. In the embedded random matrix ensemble, fermionic interactions are defined by
\begin{equation}
  H_k = \sum_{\alpha,\gamma} v_{k;\alpha,\gamma} \Psi^\dagger_{k;\alpha}\Psi_{k;\gamma},
  \label{eq:Ham}
\end{equation}
which takes into account interactions between $k$-fermions ($1\leq k\leq n$). The
$\Psi^\dagger_{k;\alpha}$ is a collective creation operator of $k$-particles. When applied to the
state $|0\rangle$ it generates a quantum state $\Psi^\dagger_{k;\alpha}|0\rangle$ of $k$-particles
distributed in $l$ levels in the specific configuration denoted by $\alpha$. For instance, if $l=6$
and $k=2$, one possible $\alpha$ configuration is
$\Psi^\dagger_{2;\alpha}|0\rangle=a^\dagger_1 a^\dagger_3|0\rangle=|1,0,1,0,0,0\rangle$, where
$a^\dagger_j$ is a fermionic creation operator. By convention, the indices of the $a^\dagger_j$ are
arranged in increasing order. The corresponding annihilation operator $\Psi_{k;\gamma}$ is
constructed analogously. The coefficients $v_{k;\alpha,\gamma}$ are independent identically
distributed Gaussian variables with zero mean and unit variance. Finally, the sum in \eq{Ham} runs
over all distinct configurations $\alpha$ and $\gamma$ of $k$-particles distributed in $l$ single
particle states.

\begin{figure}[htb]
  \centering
  \includegraphics[scale=0.26]{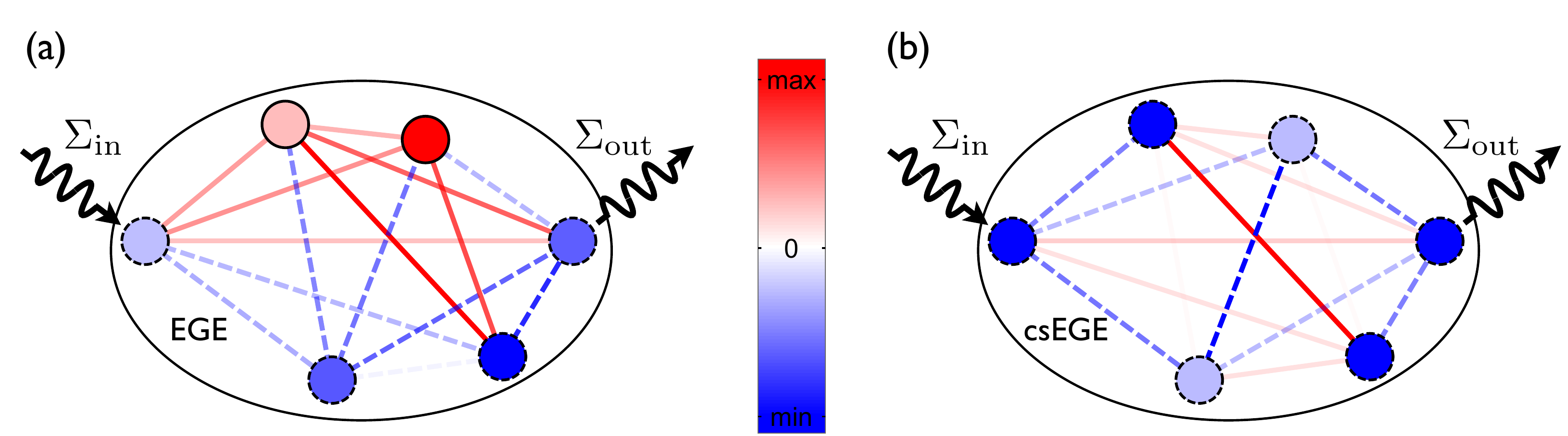}
  \caption{Network representation of Hamiltonians from the EGE (a) and csEGE (b). Excitations are
    injected ($\Sigma_{\text{in}}$) and extracted ($\Sigma_{\text{out}}$) through two specific
    states of the system. Dashed lines indicate negative values, solid lines positive values.}
  \label{fig:1}
\end{figure}

A natural basis to represent the interaction Hamiltonian $H_k$ is the occupation number basis, which
corresponds to the set $\lbrace |\mu\rangle = \Psi^\dagger_{n;\mu}|0\rangle | \mu\in S \rbrace$,
where $S$ is the set of all the possible ways in which we can distribute $n$-particles in $l$-single
particle levels. This representation of the Hamiltonian $H_k$ can be interpreted as a disordered
network, where each site represents an $n$-body many-particle state $|\mu\rangle$. The total number
of sites in the network is determined by the dimension of the Hilbert space $N=\binom{l}{n}$. A pair
of sites is coupled if the matrix element $\langle\nu|H_k|\mu \rangle\neq 0$ \cite{2003JPA-BW}. An
example for such a network is shown in \fig{1}. For the construction of the csEGEs, which is based
on preserving the centrosymmetry at the one-particle level, we refer to
Refs.~\cite{PhysRevE.94.042102, 2015AdrManBen}.

In \cite{PhysRevE.94.042102, 2015AdrManBen} we show that optimal transport properties are obtained
for both the EGE and the csEGE if the total number of particles is $n=l-1$ and $k\sim n/2$. In this
case, centrosymmetry implies \cite{OrtegaPhDThesis2017, cantoni}
\begin{equation}
  [H_k^{(cs)},J_N] = 0.
  \label{commHJ}
\end{equation}
The exchange matrix $J_N$ is defined by $J_{ij}=\delta_{i,N-j+1}$, where $\delta_{kl}$ is the
Kronecker delta. The centrosymmetric Hamiltonian $H_k^{(cs)}$ (in the occupation number basis)
attains the block structure
\begin{equation}
  H_k^{(cs)} = 
  \begin{pmatrix}
    A && C^T \\
    C && J_{N/2}AJ_{N/2}
  \end{pmatrix},
\end{equation}
where $A,J_{N/2}$ and $C$ are matrices of dimension $N/2\times N/2$ and $A=A^T$,
$C^T=J_{N/2}CJ_{N/2}$.\footnote{For concreteness, we have assumed that $N$ is even.} Using the
orthogonal transformation \cite{cantoni}
\begin{equation}
  \label{eq:1}
  \mathcal{O}=\frac{1}{\sqrt{2}}\begin{pmatrix}\mathbb{1} & -J_{N/2}\\ \mathbb{1} & J_{N/2}  \end{pmatrix},
\end{equation}
$H_k^{(cs)}$ can be cast in a block diagonal form
\begin{equation}
  \mathcal{O}H_k^{(cs)}\mathcal{O}^T=
  \begin{pmatrix}
    A-J_{N/2}C && 0 \\
    0    && A+J_{N/2}C
  \end{pmatrix}.
  \label{eq:blockstructure}
\end{equation}
Furthermore, the eigenvectors of $H^{(cs)}$ fulfill
\begin{equation}
  \begin{split}
    J_N|v\rangle & = |v\rangle, \\
    J_N|w\rangle & = -|w\rangle,
  \end{split}
  \label{eq:parity}
\end{equation}
where half of the eigenvectors are symmetric ($|v\rangle$), and the other half are skew-symmetric
($|w\rangle$). In this context, a vector in the occupation number basis obeys parity if it fulfills
either of the two equations in \eq{parity}. Therefore, using \eq{blockstructure}, we see that
$H^{(cs)}$ has parity, revealed in its block structure, and correlations between different sectors
of parity.

\subsection{Nonequilibrium Green's function method for quantum transport}
\label{sec:NEGF}

The transport of fermionic excitations in disordered networks generated from the EGE or csEGE is
studied by means of the nonequilibrium Green's function method. We summarize briefly the essential
equations. A detailed description can be found in Refs.~\cite{Datta1997, Datta2005, DiVentra2008,
  Cuevas2010}.

The Green's function of the system is defined as
\begin{equation}
  \label{eq:gf}
  G(E) = (E -H -\Sigma_{\text{in}} -\Sigma_{\text{out}})^{-1}, 
\end{equation}
where $E$ is the excitation energy and $H$ the system Hamiltonian (for example, a member of the EGE
or the csEGE). The self-energy matrix elements 
\begin{align}
  \label{eq:3}
  {\Sigma_\text{in}}_{r,s} &= - \I \eta \, \delta_{r,\text{in}} \delta_{r,s},\\
  {\Sigma_\text{out}}_{r,s} &= - \I \eta \, \delta_{r,\text{out}} \delta_{r,s},
\end{align}
describe the effect of coupling the system to an external environment (or reservoir) through which
the excitations are injected (in) and extracted (out) with rate $\propto \eta/\hbar$. Note that the
self-energies have only one nonvanishing matrix element for $r=s=\text{in/out}$. Transport is
studied between the state $\ket{\text{in}}=|1,1,\dots,1,0,0,\dots,0\rangle$, where all the fermions
are shifted to the left, and the state $\ket{\text{out}}=|0,0,\dots,0,1,1,\dots,1\rangle$, where all
the fermions are shifted to the right. Note that $\ket{\text{out}}$ is related to $\ket{\text{in}}$
by centrosymmetry, $\ket{\text{out}} = J_N \ket{\text{in}}$. We consider such centrosymmetric
related states because transport is optimal among them \cite{OrtegaPhDThesis2017}.  Considering
other pairs of centrosymmetric states yields the same results.

\begin{figure*}[htb]
  \centering
  \includegraphics[scale=0.5]{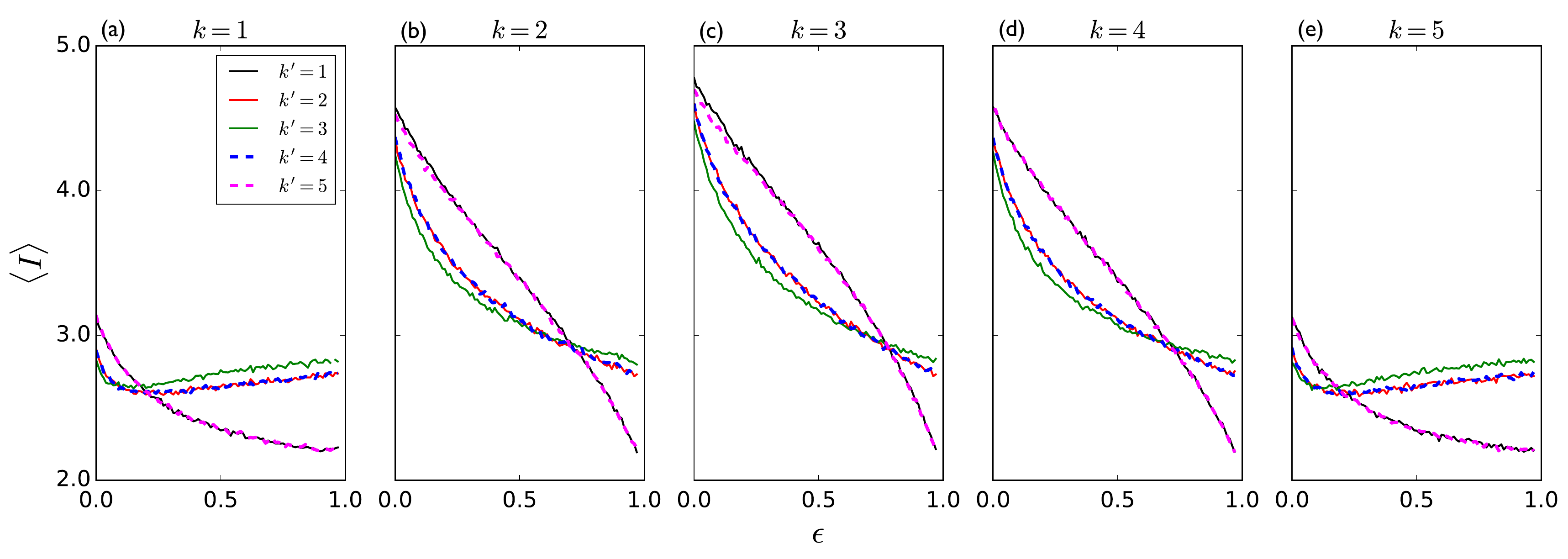}
  \vspace{-2mm}
  \caption{Perturbation of centrosymmetric Hamiltonian with a noncentrosymmetric Hamiltonian. The
    mean current $\braket{I}$ is plotted as a function of the perturbation strength $\eps$. In each
    figure the value interaction rank $k$ for the csEGE is kept constant, while the interaction rank
    $k'$ for the EGE perturbation is varied; see the inset. In the case of optimal transport ($k=3$)
    any perturbation reduces drastically the transport efficiency.}
  \label{fig:EGEtocsEGE}
\end{figure*}

A Fourier transform from the energy to the time domain shows that the matrix elements of the Green's
function $G_{i,j}(t)$ describe the response of the state $j$ at time $t$ after a $\delta(t)$
excitation of the state $i$ at time $t=0$ \cite{Datta1997, Datta2005}. Hence, the Green's function
describes the propagation of excitation through the many-body states of the quantum system. A
similar situation is found in photosynthetic complexes, where an excitation is injected at a
specific site, called the antenna, and extracted after a certain time at another specific site,
called the sink \cite{blakenship}.

The transmission probability between the states $\ket{\text{in}}$ and $\ket{\text{out}}$ is given by
\cite{Caroli1971}
\begin{equation}
  T(E) = 4 \operatorname{Tr}[\operatorname{Im}(\Sigma_{\text{in}})G\operatorname{Im}(\Sigma_{\text{out}})G^\dagger].
  \label{eq:T}
\end{equation}
The ensemble-averaged total current, which can be driven through the system, is given by \cite{PhysRevE.94.042102}
\begin{equation}
  \langle I \rangle = \left\langle \int_{-\infty}^\infty \text{d}E \, T(E) \right\rangle.
  \label{eq:I}
\end{equation}
This quantity will be used below to benchmark the efficiency of quantum transport in the system.

\section{Results and discussion}
\label{sec:ResultsDiscussion}

In this Section, we present the results of perturbing the transport in disordered interacting
systems. In our previous work \cite{2015AdrManBen, PhysRevE.94.042102}, we have shown that optimal
transport is obtained in a system of $l$ states if these states are occupied with $n=l-1$ fermions
interacting via $(k\sim n/2)$-body interactions. Hence, we will focus our investigations on the
optimal case $(l,n,k)=(6,5,3)$. Unless otherwise stated, all ensembles comprise $10^4$ realizations.

\subsection{Mixture of csEGE and EGE}
\label{sec:EGEtocsEGE}

\begin{figure*}[htb]
  \centering
  \includegraphics[scale=0.5]{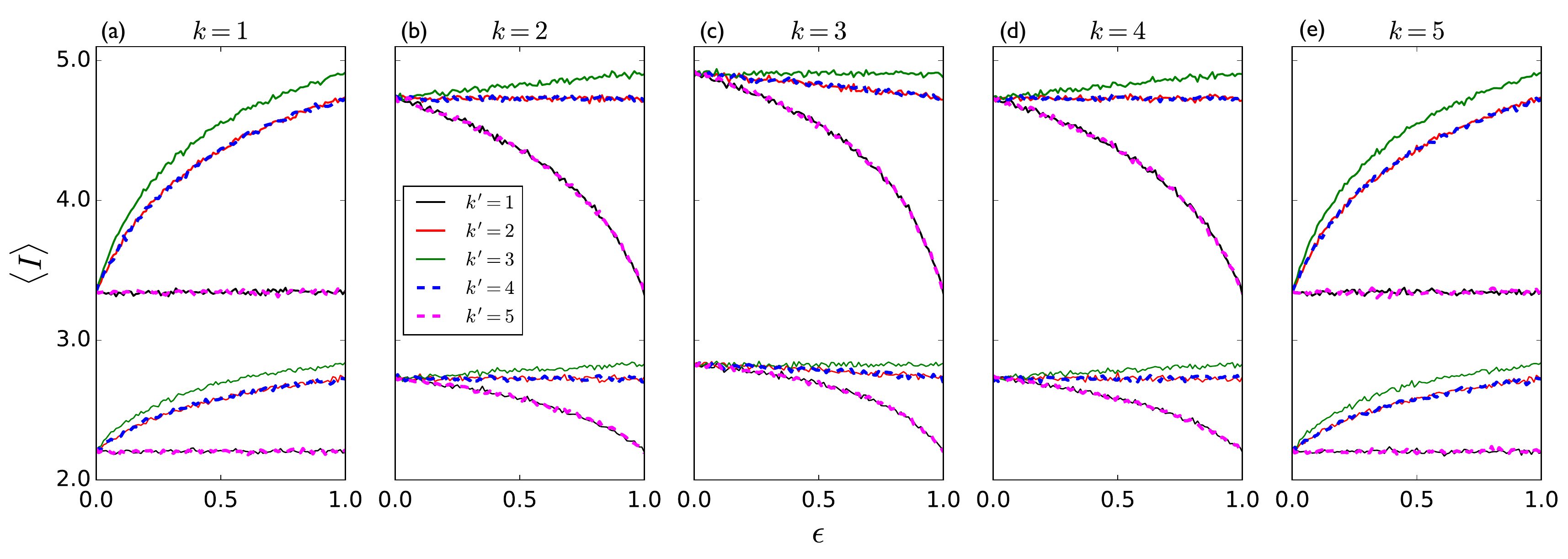}
  \vspace{-3mm}
  \caption{Mixing two Hamiltonians from the csEGE (top curves) and from the EGE (bottom curves). The
    systems pass between the cases of optimal transport ($k \sim 3$) and suboptimal transport
    ($k =1,5$). In general, the transport in the csEGE is much more efficient than in the EGE.}
  \label{fig:EGEtoEGE}
\end{figure*}

We add to a centrosymmetric Hamiltonian $H_k^{(cs)}$ a noncentrosymmetric perturbation
$H_{k'}$ by means of the model
\vspace{-3mm}
\begin{equation}
  H_T = \sqrt{1-\eps}H_k^{(cs)} + \sqrt{\eps}H_{k'},
  \label{eq:EGEtocsEGE}
\end{equation}
where $\eps \in [0,1]$ controls the strength of the perturbation. Both Hamiltonians have the same
values for $l$ and $n$. Such a mixture of ensembles is a paradigmatic case, because in many
situations the system is composed of one- and two-body interactions \cite{2010PRE-MananKota,
  2009PLA-MananKota}. However, here we investigate a much broader parameter space, because $k$ can
vary between 1 and $n$. Note that the perturbation strength is scaled by a square root in order to
keep the spectral span and the current constant in the case that both Hamiltonians are from the same
ensemble with the same $k=k'$, see for example the horizontal curves in \fig{EGEtoEGE}. In the
subsequent figures $\epsilon$ is varies from $0$ to $1$ with steps of size
$\Delta \epsilon= 10^{-2}$. For each value of $\epsilon$ we calculate the corresponding ensemble
average $\braket{I}$.

The result of this perturbation is shown in \fig{EGEtocsEGE}. The ensemble-averaged total current
$\braket{I}$ is plotted as a function of the parameter $\eps$. We observe that in the case of
optimal transport ($k \sim 3$) any perturbation reduces drastically and rapidly the transport. In
particular, a weak perturbation ($\eps=0.25$) with $k'=2,3,4$ reduces the total current by
approximately $30 \%$. On the other hand, in the case of suboptimal transport ($k=1,5$) the effect
of the perturbation is much weaker. Note that the case $k=1$ can also be interpreted as lifting the
degeneracy of the single-particle states. We observe that perturbing with $k'=2,3,4$ initially
degrades the current (because centrosymmetry is broken), while stronger perturbations again enhance
the transport (because the transport is generally better for $k' =2,3,4$). Note that
\fig{EGEtocsEGE} also confirms our previous findings \cite{PhysRevE.94.042102} that the transport
efficiency in centrosymmetric systems ($\eps=0$) is higher than in noncentrosymmetric systems
($\eps=1$). The total current as a function of the two parameters $k$ and $k'$ shows several
symmetries
$\braket{I(k,k')}= \braket{I(n{-}k,k')}= \braket{I(k, n{-}k')}= \braket{I(n{-}k,n{-}k')}$. This is a
consequence of the way in which the ensemble is defined [see \eq{Ham}] and has nothing to do with
particle-hole symmetry \cite{2003JPA-BW, PhysRevE.94.042102}. These symmetries will also appear in
the perturbations discussed below.

When two Hamiltonians from the EGE or from the csEGE are mixed, as shown in \fig{EGEtoEGE}, we
observe a transition between the cases of optimal transport ($k \sim 3$) and suboptimal transport
($k=1,5$). Moreover, it is confirmed clearly that centrosymmetry (top curves) enhances significantly
the transport efficiency compared to noncentrosymmetric systems (bottom curves).

\subsection{Breaking parity and centrosymmetry by block perturbations}
\label{sec:csEGEbreaking}

Taking into account the block diagonal form of the Hamiltonian \eq{blockstructure}, the matrix that
breaks parity can be written as
\begin{equation}
  \mathcal{O}H_B\mathcal{O}^T = 
  \begin{pmatrix}
    0 && B \\
    B && 0
  \end{pmatrix},
\end{equation}
where for simplicity we consider $B$ as a member of the Gaussian orthogonal ensemble. It is evident
that the off-diagonal blocks mix different parity sectors in \eq{blockstructure}. In the occupation
number basis, this perturbation takes the form
\begin{equation}
  H_B = 
  \begin{pmatrix}
    B && 0 \\
    0 && -J_{N/2}BJ_{N/2}
  \end{pmatrix},
  \label{eq:Bparity}
\end{equation}
and thus, parity breaking in the basis, where $H$ has block structure, is equivalent to a diagonal
perturbation by blocks in the occupation number basis of $n$ particles. We model parity breaking by 
\begin{equation}
  H_T = \sqrt{1-\eps}H^{(cs)}_k +\sqrt{\eps}H_B.
  \label{eq:perturbationB}
\end{equation}

\begin{figure}[htb]
  \centering
  \includegraphics[scale=0.45]{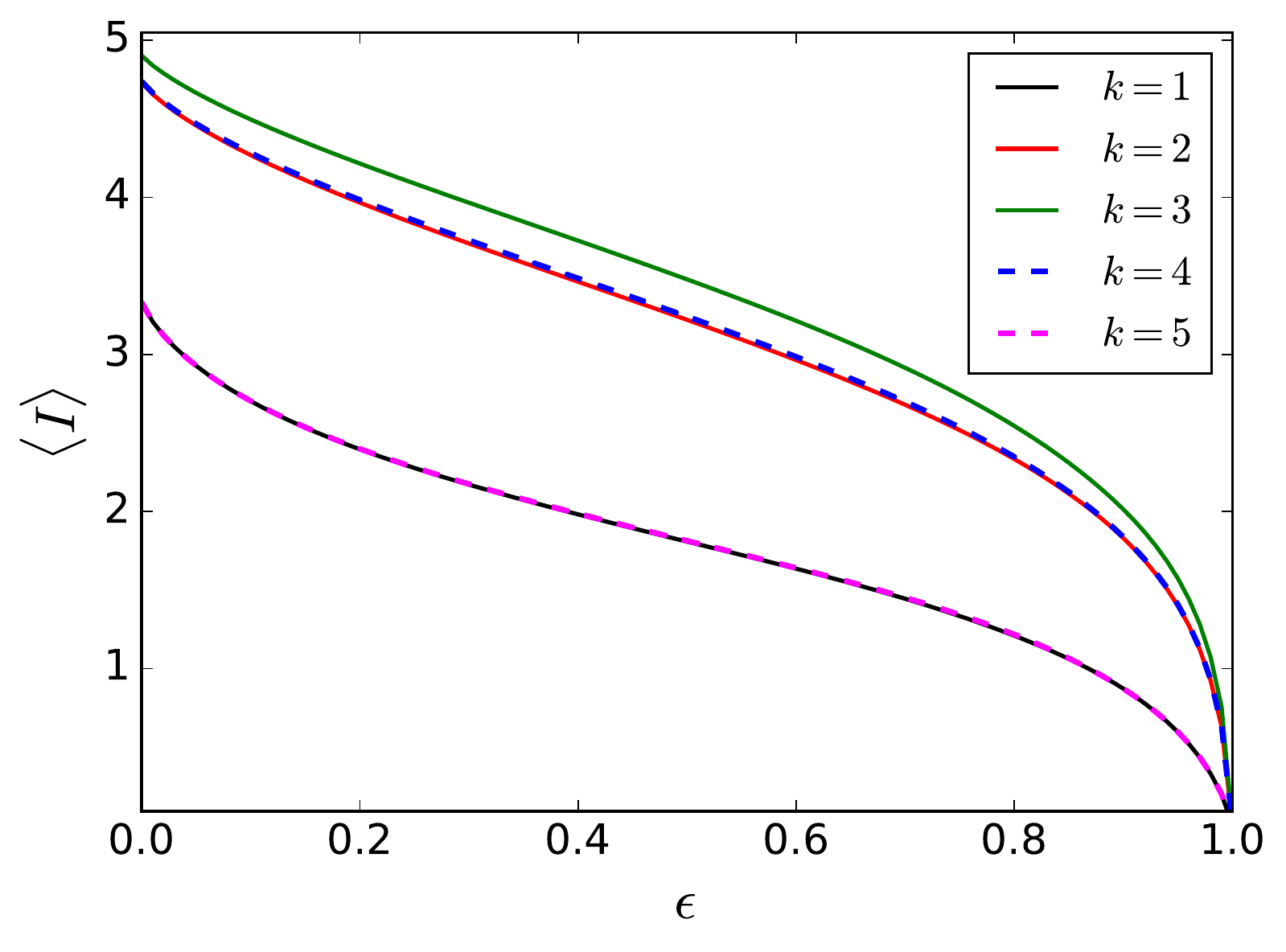}
  \vspace{-2mm}
  \caption{Effect of parity breaking on the total current in a centrosymmetric system. The different
    curves represent csEGEs with different values of $k$; see the inset. For $\eps=0$ we obtain the
    average total current $\braket{I}$ for the csEGE. For $\eps=1$ the system consists of two
    independent block and transport is blocked completely ($\braket{I}=0$).}
  \label{fig:paritybreaking}
\end{figure}

\Fig{paritybreaking} shows the effect of parity breaking on the current $\braket{I}$ as a function
of $\eps$. For all values of $k$, parity breaking reduces significantly the current. In the case of
optimal transport ($k\sim 3$), the current decays approximately linearly (after a very short
seemingly quadratic decay) and approaches for $\eps \approx 0.8$ the corresponding values of the
EGE.  For $k=1,5$ this values is obtained already for $\eps \approx 0.2$. For $\eps \to 1$ a strong
reduction of the current is observed, because the Hamiltonian of the system $H_T= H_B$ consists of
two independent blocks. As the injecting and extracting reservoirs ($\ket{\text{in}}$ and
$\ket{\text{out}}$) are located in different blocks, the two reservoirs are effectively decoupled
and transport gets completely suppressed ($\braket{I}=0$).

Another perturbation is the breaking of centrosymmetry, modeled by
\begin{equation}
  H_T = \sqrt{1-\eps}H^{(cs)} + \sqrt{\eps}H_D,
  \label{eq:pertbyblocks}
\end{equation}
where $H_D$ (in the occupation number basis) is defined as
\begin{equation}
  H_D=\begin{pmatrix}
    0 && D \\
    D^T && 0
  \end{pmatrix}.
  \label{eq:Dperturbation}
\end{equation}
$D$ is a real square matrix with Gaussian normal variables in each entry and hence, generally not
symmetric.

\begin{figure}[htb]
  \centering
  \includegraphics[scale=0.45]{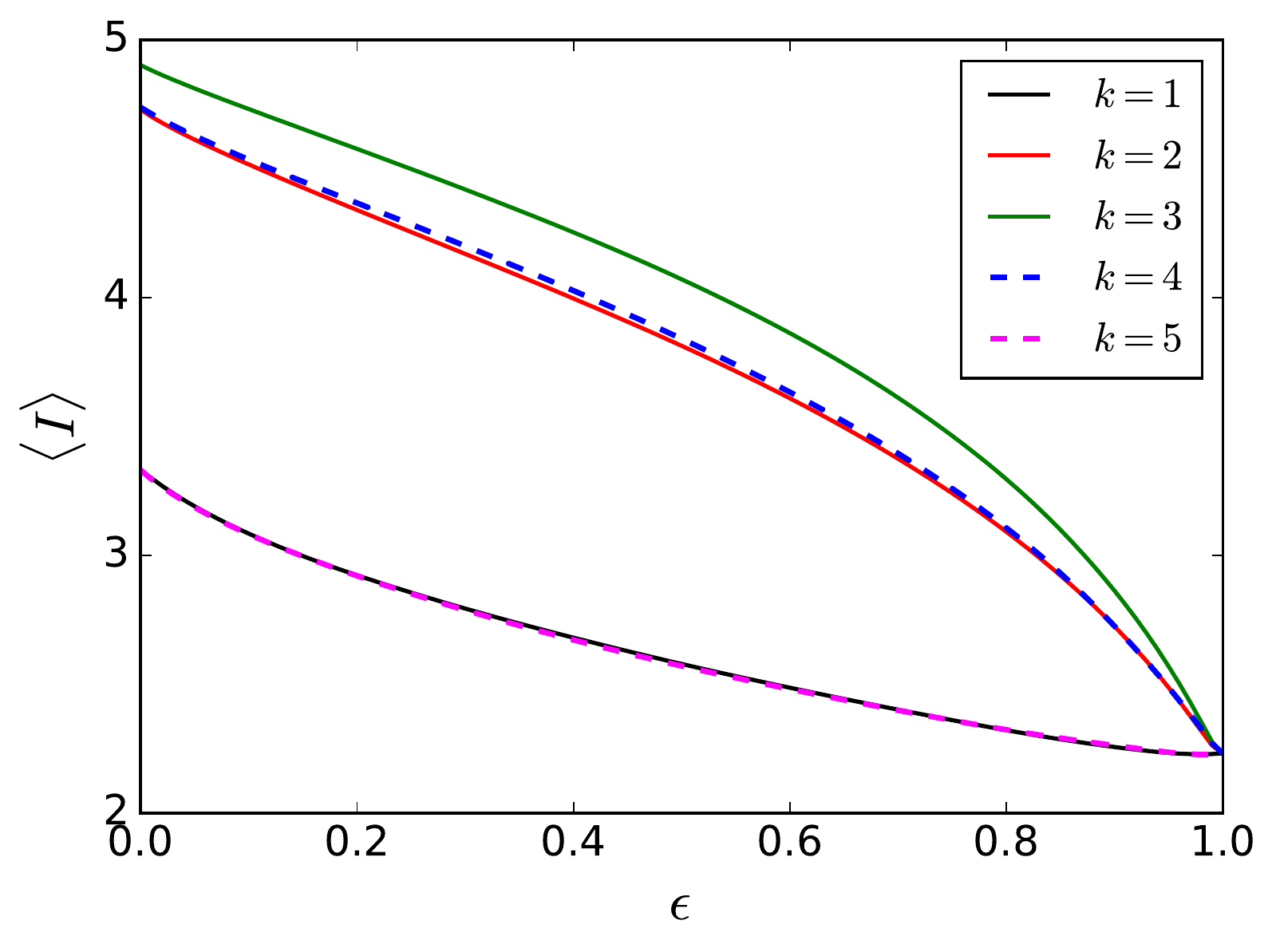}
  \vspace{-2mm}
  \caption{Effect of centrosymmetry breaking effect on the total current in a centrosymmetric
    system. For all $k$ the current decreases continuously. The minimal value for $\eps=1$ is
    independent of $k$ because in this case the system consists of Gaussian random matrices.}
  \label{fig:pertbyblocks}
\end{figure}

\Fig{pertbyblocks} shows the effect of centrosymmetry breaking by applying the off-diagonal block
perturbation with the matrix $D$. For all $k$ the current decreases until its final value. In the
case of optimal transport ($k \sim 3$) the current decreases $ \sim 50 \%$, while for $k=1,5$ it
decreases only $\sim 33 \%$.

\subsection{Contact influence in coherent transport}
\label{sec:contactinfluence}

\begin{figure}[htb]
  \centering
  \includegraphics[scale=0.35]{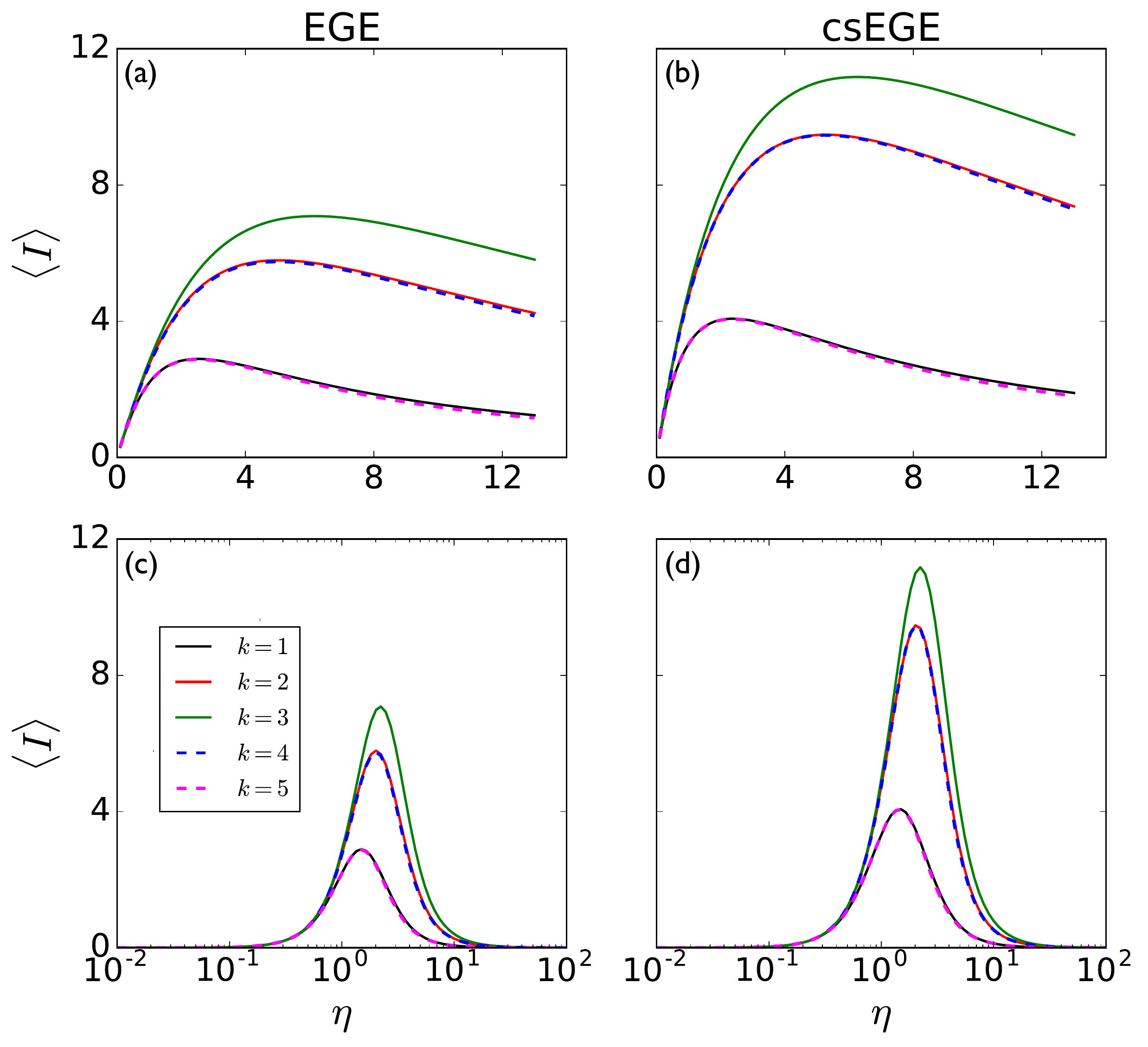}
  \vspace{-2mm}
  \caption{Current $\braket{I}$ as a function of the coupling parameter $\eta$. The left and the
    right columns are for the EGE and csEGE, respectively. Note that the top row the scale of $\eta$
    is linear, while in the bottom row it is logarithmic. The current is maximal for a specific
    value of $\eta$.}
  \label{fig:eta}
\end{figure}

A parameter that is often ignored in studies of quantum efficiency is the influence of the coupling
of the environment (or reservoirs) to the central system. Hence, we analyze in \fig{eta} how the
current is affected by the parameter $\eta$ [see \eq{3}], which parametrizes the coupling strength
of the two real reservoirs. The left column indicates the case of the EGE and the right column for
the csEGE. In the top row the scale of $\eta$ is linear, while in the bottom row it is logarithmic.

The main result is that the current is maximal for a specific finite value of $\eta$, which depends
on $k$. The current decreases if the coupling is weakened or if the coupling gets too
strong. Comparing the two columns, it can be observed that for a fixed value of $\eta$ the csEGE
provides a higher current than the EGE. Note that the described properties are similar to the
superradiance transition \cite{celardo}. The transport can be understood also in terms of the
transfer time through the system. Hence, we can interpret the optimal coupling strength as the one
where the transfer time matches the rate at which the excitations are injected and extracted.

\subsection{Transport in the EGE and csEGE in presence of decoherence}
\label{sec:deph}

\begin{figure*}[htb]
  \centering
  \includegraphics[scale=0.45]{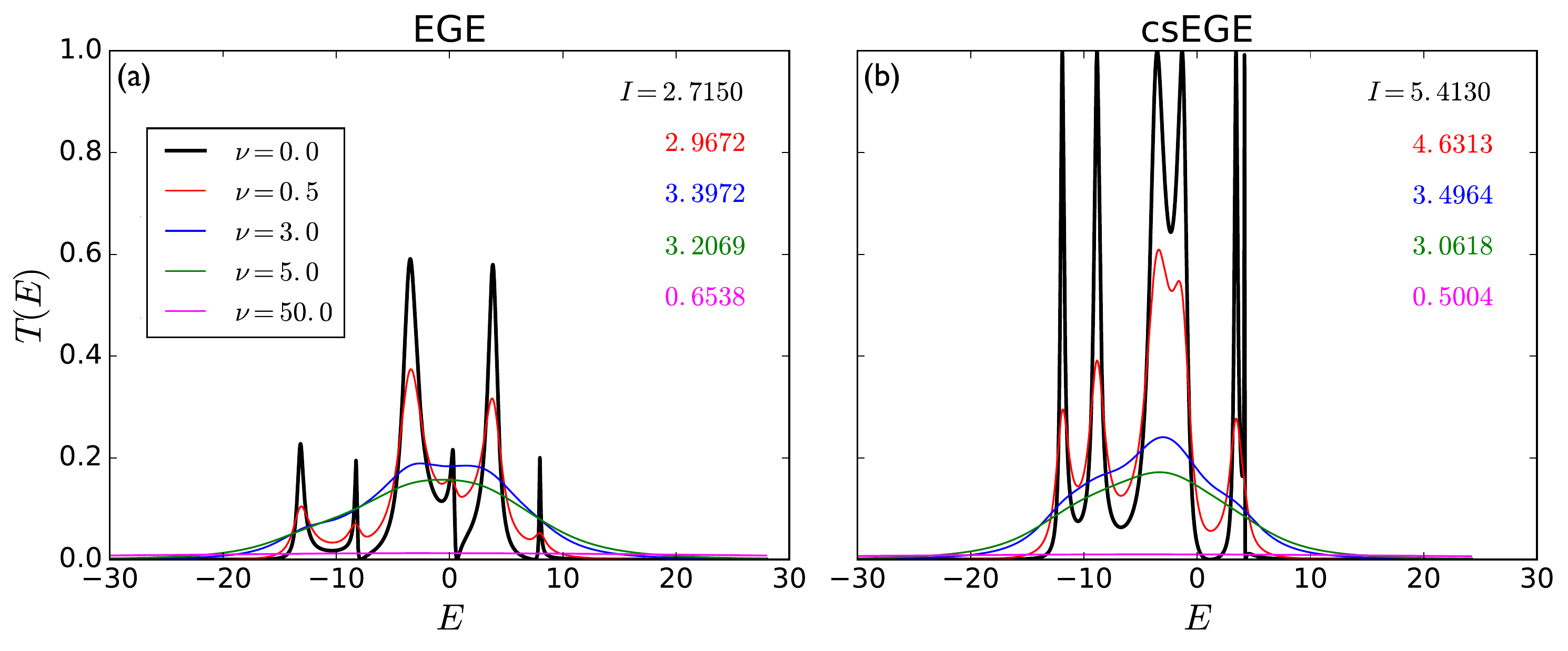}
  \vspace{-2mm}
  \caption{Transmission $T(E)$ for a realization from the EGE (a) and csEGE (b) under the effect of
    decoherence. The decoherence strength is controlled by the parameter $\nu$. The total current,
    given by the area below the curves, is indicated in the inset. In general, decoherence smooths
    out the transmission resonances. In the EGE, the current attains its maximum for a finite
    nonzero value of $\nu$ (decoherence assisted transport). In the csEGE, the decoherence always
    reduces the current.}
  \label{fig:deco_onerealiz}
\end{figure*}

\begin{figure*}[htb]
  \centering
  \includegraphics[scale=0.55]{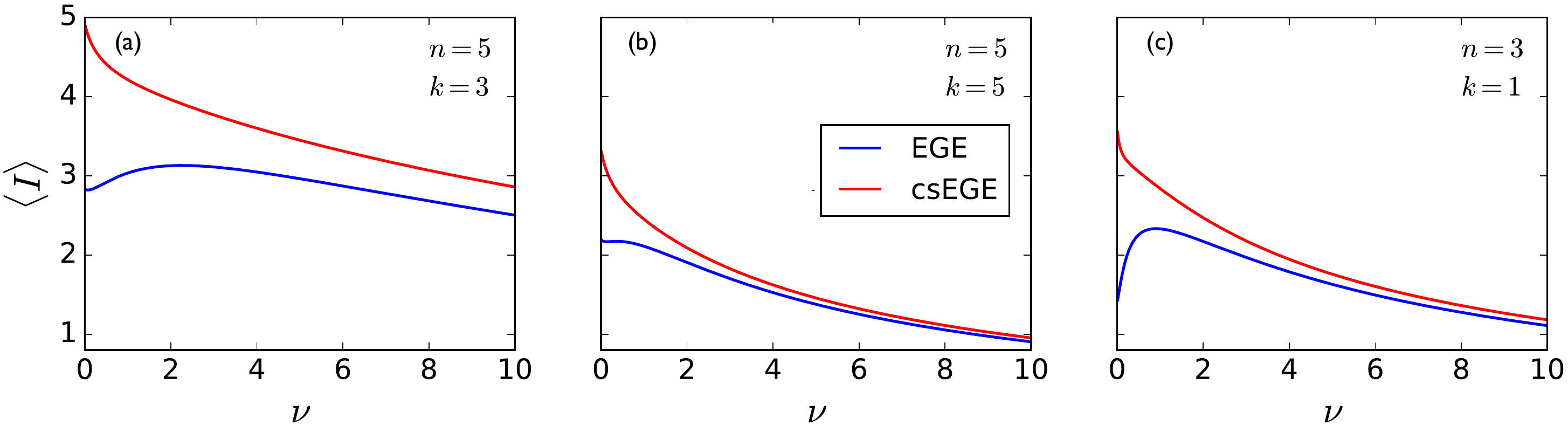}
  \vspace{-2mm}
  \caption{Total current $\braket{I}$ as a function of the decoherence strength $\nu$. In all cases
    $l=6$, while the values of $(n,k)$ are indicated in the inset of each figure. (a) corresponds to
    parameters of optimal transport, (b) corresponds to a completely uncorrelated disordered network
    and, (c) corresponds to a many-particle system with one-body interactions. In the EGE (blue
    curves), the current approaches its maximum at a finite nonzero value of $\nu$, which is known
    as environment-assisted transport. In the csEGE (red curves), the current decays monotonically
    with increasing decoherence strength but is always larger than in the EGE. Hence, the
    correlations induced by centrosymmetry enhance the transport much more than decoherence, which
    suppresses Anderson localization but also breaks correlations.}
  \label{fig:deco_Iaverage}
\end{figure*}

We study the effects of decoherence on the transport efficiency, comparing in particular its
interplay with centrosymmetry. In order to take into account the effects of decoherence, we use
B\"uttiker's approach of fictitious reservoirs, where excitations are absorbed and re-injected after
randomization of their phase \cite{PhysRevB.33.3020}. This idea has been generalized by Pastawski to
a continuous distribution of fictitious probes \cite{Amato1990}. Following this work we attach to
each state $|\mu\rangle$ a fictitious reservoir that is modeled by the self-energy
\begin{equation}
  \label{eq:nu}
  {\Sigma_\mu}_{r,s} = - \I \nu \, \delta_{r,\mu} \delta_{r,s},
\end{equation}
which also have to be taken into account in the Green's function \eq{gf}. The coupling strength
$\nu$ of the virtual reservoirs determines the decoherence strength. The system now comprises of two
real reservoirs (in, out), through which the excitations are injected and extracted, as well as $N$
virtual reservoirs, which model the effects of decoherence. The transmission through the system is
now given by the D'Amato-Pastaswki model \cite{Amato1990, StegmannPhD}
\begin{equation}
  T(E) = T_{\text{in},\text{out}} + \sum_{ij}T_{\text{in},i}\,\mathcal{R}_{ij}\,T_{j,\text{out}},
  \label{eq:Imulti}
\end{equation}
where 
\begin{equation*}
  \mathcal{R}_{ij}^{-1} = \left\{
  \begin{array}{ll}
    -T_{ij}, & i\neq j,\\[2mm]
    \sum_{k\neq i} T_{ik}, & i=j.
  \end{array}
  \right.
\end{equation*}
The transmission $T_{ij}$ from reservoir $i$ to reservoir $j$ can be calculated by means of \eq{T}
using the corresponding self-energies.

\Fig{deco_onerealiz} shows the transmission $T(E)$ for a typical realization from the EGE (a) and
csEGE (b). In the case of coherent transport ($\nu=0$), centrosymmetry generates several resonances
of perfect transmission ($T(E)=1$); see our previous work \cite{PhysRevE.94.042102} for
details. When the decoherence increases, the transmission resonances are smoothed out. The total
current $\braket{I}$ as a function of the decoherence strength $\nu$ is shown for various system
parameters $(n,k)$ in \fig{deco_Iaverage}. In the EGE (blue curves), the current obtains its maximum
for a finite nonzero value of $\nu$. This is the decoherence assisted transport
\cite{plenio_dephtransp, rebentrost_dephtransp}, where the loss of the height of the resonance peaks
is overcompensated by the broadening of the resonances (c.f. \fig{deco_onerealiz}) and hence, the
environment fosters transport. In the csEGE (red curves), the total current decreases monotonically
under the effect of decoherence. In the same way as decoherence suppresses Anderson localization and
fosters transport, in the present case it also destroys the correlations induced by
centrosymmetry. In spite of this, it can be observed clearly that centrosymmetry enhances transport
much more than decoherence, as manifested by the total current, which in the csEGE is always larger
than in the EGE. Finally, for strong decoherence ($\nu=50$) the transport is completely blocked in
both ensembles. We emphasize that these results apply to the ensemble-averaged current. For specific
values of the energy, as can be read in \fig{deco_onerealiz}, the transmission can be enhanced or
decreased by decoherence depending on the actual value of the energy considered. In particular,
close to a resonance we observe that decoherence may increase the transmission, while far from it
the transmission is suppressed. This behavior was noticed already by D'Amato and Pastawski
\cite[cf. Fig 3]{Amato1990} for a certain noncentrosymmetric system.

\section{Conclusions}
\label{sec:Conclusions}

We have studied the robustness of the transport efficiency in disordered interacting many-body
quantum systems, addressing in particular the role of centrosymmetry. The efficiency has been
quantified by the average total current $\braket{I}$ that can be driven through the system.

We have analyzed how the transport efficiency is affected when a centrosymmetric $k$-body EGE is
perturbed by a $k'$-body noncentrosymmetric one, see \fig{EGEtocsEGE}. It was found that in the
optimal cases ($k \sim 3$) the efficiency is reduced significantly, while in the suboptimal cases
the efficiency is less affected. When two Hamiltonians from the csEGE or from the EGE are mixed [see
\fig{EGEtoEGE}] it is clearly observed that the transport in centrosymmetric systems is always
better than in the corresponding noncentrosymmetric systems.  We have studied the effect of block
perturbations that break parity and centrosymmetry, see \fig{paritybreaking} and
\fig{pertbyblocks}. It was found that, similarly to the case of mixing csEGE with EGE, the transport
efficiency decays to a minimal value. Investigating the effect of the coupling strength $\eta$ to
the environment, we have shown in \fig{eta} that the transport efficiency approaches a maximum at a
specific value of $\eta$, whereas too weak and too strong coupling hinders the transport. Finally,
analyzing the interplay of decoherence and centrosymmetry in \fig{deco_onerealiz} and
\fig{deco_Iaverage}, we have found that in the EGE the transport efficiency can be enhanced by
decoherence, which is known as environment-assisted transport. In the csEGE the efficiency is
reduced monotonically by decoherence, and therefore there is no signature of environment-assisted
transport. We interpret such suppression of transport as a consequence of decoherence affecting the
correlations induced by centrosymmetry and parity. Yet, the resulting net current is always higher
than for the noncentrosymmetric ensemble.

The results about decoherence are interesting in various aspects. While it is not clear if
centrosymmetry is present in efficient photosynthetic biomolecules, it certainly defines an
alternative for the design of efficient transport devices. The resulting transport properties in
presence of centrosymmetry are an improvement over those by environment-assisted transport, and may
likely exceed also the superradiance controlled by disorder \cite{Doria2018}. These results could be
experimentally tested using finite discrete optical lattices or in spin chains with NMR techniques;
see \cite{nikojex}. Thus, centrosymmetry represents a valuable option worth considering for optimal
transport.

\vspace*{5mm}

\begin{acknowledgments}
  We acknowledge financial support from the projects UNAM-PAPIIT IG-100616 and IA-101618, as well as
  project CONACYT Fronteras 952.
\end{acknowledgments}

\bibliographystyle{apsrev4-1}
\bibliography{bibliografia.bib}

\end{document}